\documentclass[aps,
pre,
reprint,
superscriptaddress,longbibliography]{revtex4-2}

\usepackage{amsmath,amssymb,amsfonts,bm}
\usepackage{graphicx}
\usepackage{hyperref}

\newcommand{\ee}{\mathrm{e}}
\newcommand{\ii}{\mathrm{i}}

\newcommand{\ip}[2]{\left\langle #1, #2 \right\rangle_{\pi}}
\newcommand{\normpi}[1]{\left\lVert #1 \right\rVert_{2,\pi}}
\newcommand{\norminf}[1]{\left\lVert #1 \right\rVert_{\infty}}
\newcommand{\Dform}{\mathcal{D}}
\newcommand{\EP}{\sigma}

\begin{document}

\title{Dissipation-coherence tradeoff for stochastic oscillations}

\author{Jie Gu}
\affiliation{Chengdu Academy of Educational Sciences, Chengdu 610036, China}

\date{\today}

\begin{abstract}
Autonomous noisy oscillations in biochemical and mesoscopic systems require nonequilibrium driving and therefore dissipation. A striking conjecture by Oberreiter, Barato, and Seifert (OBS) proposes a universal lower bound on the entropy produced per oscillation period in terms of the coherence number of the slowest oscillatory mode. Here we derive a weaker but rigorous lower bound that preserves the OBS structure while introducing a mode-uniformity factor that quantifies how evenly the oscillatory eigenmode is distributed across states in the steady-state inner product. The result makes explicit that an eigenvalue-only prefactor can fail when the dominant oscillatory mode is localized. We also outline a proof-of-principle route for estimating this factor from low-dimensional data under single-mode dominance and sufficiently informative measurements, and derive an eigenvector-free corollary using only the smallest stationary probability. Translation-invariant Markov jump processes on a ring provide a symmetry-protected class with $\eta=1$, so the refinement reduces to the OBS form; the drift--diffusion limit on a circle saturates the bound.
\end{abstract}

\maketitle


\section{Introduction}

Many physical and biological systems exhibit noisy oscillations without being driven by an external periodic signal. Examples include biochemical clocks, enzymatic cycles, and mesoscopic reaction networks \cite{Goldbeter1996,ElowitzLeibler2000,Gillespie1977}. Unlike deterministic limit-cycle oscillators, a finite stochastic system cannot oscillate coherently forever: phase diffusion and amplitude decay limit the number of visible cycles \cite{VanKampen2007,Gardiner2009,BaratoSeifert2017,OBS2022}. Understanding the thermodynamic price of maintaining such oscillations is therefore a basic question in nonequilibrium statistical physics and in the physics of biochemical clocks \cite{Seifert2012,BaratoSeifert2017,OBS2022}.

Several lines of work have clarified this question from complementary viewpoints. In biochemical oscillator models, free-energy consumption has been related to phase accuracy, phase sensitivity, synchronization performance, and design under limited energetic resources \cite{CaoWangOuyangTu2015,FeiCaoOuyangTu2018,ZhangCaoOuyangTu2020,CaoJiangHou2020}. At a broader level, stochastic thermodynamics has supplied a general framework for open chemical reaction networks and their dissipation \cite{Seifert2012,RaoEsposito2016}. For Markov-state descriptions, Barato and Seifert showed that coherence is constrained by network topology and driving \cite{BaratoSeifert2017}, and Oberreiter, Barato, and Seifert (OBS) conjectured a particularly striking eigenvalue-based lower bound on the entropy produced per oscillation period,
\begin{equation}
\Delta S \ge 4\pi^2 \mathcal{N},
\label{eq:OBS}
\end{equation}
where $\mathcal{N}$ is the coherence number of the slowest oscillatory mode, provided $\mathcal{N}\ge (2\pi)^{-1}$ \cite{OBS2022}. Related results bound oscillatory signatures through cross-correlation asymmetry, fluctuation observables, or spectral comparison with reference dynamics \cite{MarslandCuiHorowitz2019,OhgaItoKolchinsky2023,LiangPigolotti2023,Shiraishi2023,KolchinskyOhgaIto2024,Gu2024}. Most of these results are naturally read as upper bounds on coherence at fixed thermodynamic driving, or as spectral constraints formulated with additional structure.

The question addressed here is more specific: what lower bound on dissipation per oscillation period can be proved in full generality while retaining the OBS structure? The answer turns out to depend not only on the dominant oscillatory eigenvalue but also on the geometry of its eigenvector. After introducing the Markov-process setup in Sec.~II, we prove a rigorous OBS-shaped inequality with an additional factor $\eta\in(0,1]$ that measures how uniformly the dominant oscillatory mode is distributed in the steady-state inner product. This localization correction clarifies when an eigenvalue-only universal prefactor can fail, and also identifies a symmetry-protected delocalized class for which the correction disappears. We further derive an eigenvector-free corollary based on the smallest stationary probability.

The manuscript is organized as follows. Section~II states the theorem and illustrates it numerically with three ensembles chosen to disentangle symmetry-protected delocalization, disorder-induced localization, and the role of dense connectivity. Section~III then presents a proof-of-principle inference route for $\eta$ from low-dimensional data. That section is deliberately modest in scope: it gives a heuristic construction under single-mode dominance and sufficiently informative measurements rather than a general recovery theorem. Section~IV derives the eigenvector-free corollary, Sec.~V explains why translation invariance implies $\eta=1$ and therefore recovers the OBS prefactor, Sec.~VI relates the result to existing thermodynamic bounds on oscillations, and Sec.~VII concludes.

\section{Main result}
\subsection{Setup, notation and main result}
Consider a finite irreducible continuous-time Markov jump process on states $i\in\{1,\dots,n\}$ with transition rates $k_{ij}\ge 0$ ($i\neq j$) \cite{VanKampen2007,Gardiner2009,Norris1997}. Here $k_{ij}$ denotes the jump rate $i\to j$.
Assume bidirectionality on each undirected edge: $k_{ij}>0$ if and only if $k_{ji}>0$.
Let $P(t)=(P_1(t),\dots,P_n(t))^{\top}$ be the probability column vector.
The dynamics is described by the  master equation
\begin{equation}
\dot P(t)=W P(t),
\qquad
W_{ij}=
\begin{cases}
k_{ji}, & i\neq j,\\[2pt]
-\sum_{\ell\neq i}k_{i\ell}, & i=j,
\end{cases}
\label{eq:generator}
\end{equation}
so that $W_{ij}$ is the transition rate from state $j$ to state $i$.
Irreducibility implies a unique stationary distribution $\pi=(\pi_1,\dots,\pi_n)$ with $\pi_i>0$, $\sum_i\pi_i=1$, satisfying $W\pi=0$.
It is convenient to also introduce the backward generator $L:=W^{\top}$ acting on state functions $f$, i.e.\ $(Lf)_i=\sum_{j\neq i}k_{ij}\bigl(f_j-f_i\bigr)$.

Define stationary edge flows $a_{ij}:=\pi_i k_{ij}$, edge currents $J_{ij}:=a_{ij}-a_{ji}$, and symmetric edge traffic $T_{ij}:=a_{ij}+a_{ji}$.
The steady-state entropy production rate is \cite{Schnakenberg1976,LebowitzSpohn1999,Seifert2012}
\begin{equation}
\EP = \sum_{i<j} J_{ij} A_{ij},\qquad A_{ij}:=\ln\frac{a_{ij}}{a_{ji}}.
\label{eq:sigma_def}
\end{equation}


Let the dominant oscillatory eigenvalue pair (real part closest to $0$) of $L$ be
\begin{equation}
\lambda_{1,2} = -\lambda_R \pm \ii \lambda_I, \qquad \lambda_R>0,\ \lambda_I>0,
\label{eq:lambda_pair}
\end{equation}
with corresponding right eigenvector $v\in\mathbb{C}^n$ satisfying
\begin{equation}
Lv = \lambda v,\qquad \lambda=-\lambda_R+\ii\lambda_I,\qquad v\neq 0.
\label{eq:eigenpair}
\end{equation}
The eigenvalue $\lambda=-\lambda_R+\ii\lambda_I$ controls both the oscillation frequency $\lambda_I$ and the decay rate $\lambda_R$ of the dominant oscillatory component \cite{VanKampen2007,Gardiner2009}.
Following \cite{OBS2022,BaratoSeifert2017}, define the coherence number
\begin{equation}
\mathcal{N} := \frac{1}{2\pi}\frac{\lambda_I}{\lambda_R}.
\label{eq:coherenceN}
\end{equation}
The entropy produced per oscillation period is
\begin{equation}
\Delta S := \sigma\,\frac{2\pi}{\lambda_I}.
\label{eq:DeltaS_def}
\end{equation}


Introduce the $\pi$-weighted inner product \cite{LevinPeresWilmer2009}
\begin{equation}
\ip{u}{w} := \sum_{i=1}^n \pi_i\, u_i^* w_i,
\label{eq:inner}
\end{equation}
and the induced norm $\normpi{v}^2:=\ip{v}{v}$.
Let $\norminf{v}^2 := \max_i |v_i|^2$.
Define the mode-uniformity factor by
\begin{equation}
\eta := \frac{\normpi{v}^2}{\norminf{v}^2}\in(0,1].
\label{eq:eta_def}
\end{equation}
Intuitively, $\eta$ is close to one when the eigenmode is delocalized over states of nonnegligible stationary weight, and small when the eigenmode concentrates on a small stationary subset.

Now we present our main result: the steady-state entropy production rate satisfies
\begin{equation}
\sigma \ge \eta\,\frac{\lambda_I^2}{\lambda_R}.
\label{eq:MainSigma}
\end{equation}
Equivalently, the entropy produced per period obeys the OBS-shaped inequality
\begin{equation}
\Delta S \ge 4\pi^{2}\eta\,\mathcal{N}.
\label{eq:MainDeltaS_proved}
\end{equation}
See Appendix~\ref{app:proof} for derivation.
The physical meaning of this bound parallels the OBS conjecture \cite{OBS2022}: a stochastic system cannot sustain a large spectral coherence unless it dissipates a minimum amount of free energy. The difference is that the required dissipation is modulated by the mode-uniformity factor $\eta$. If the oscillatory eigenmode is delocalized over states that carry substantial stationary weight, then $\normpi{v}^2$ and $\norminf{v}^2$ are comparable and $\eta\approx 1$, so Eq.~\eqref{eq:MainDeltaS_proved} approaches the conjectured prefactor $4\pi^2$. If the mode is localized, $\norminf{v}^2$ is large compared to $\normpi{v}^2$, so $\eta\ll 1$ and the minimum dissipation compatible with a given $\mathcal{N}$ can be parametrically smaller. As shown in Sec.~\ref{sect:eta1}, translation-invariant rings provide a concrete symmetry-protected case with $\eta=1$.

Unlike the OBS conjecture, the rigorous bound does not require the condition $\mathcal{N}\ge (2\pi)^{-1}$. This absence of a threshold should, however, be interpreted carefully. When $\mathcal{N}$ is of order unity or larger, the oscillatory component survives for many periods and $\Delta S$ has the direct meaning of entropy produced per dynamically resolved cycle. When $\mathcal{N}\ll 1$, the leading eigenvalue pair is still complex and the timescale $2\pi/\lambda_I$ remains well defined, but the oscillatory component decays before completing a clearly resolved cycle. In that weakly oscillatory regime, Eq.~\eqref{eq:MainDeltaS_proved} is therefore best read primarily as a spectral dissipation bound involving the ratio $\lambda_I/\lambda_R$, rather than as a statement about experimentally countable cycles. The operational significance of Eq.~\eqref{eq:MainSigma} is that the sharpness of the bound is governed by $\eta$, which can sometimes be approximated from partial or coarse observations.

\subsection{Example}
\label{sec:numerical_example}

We verify the localization-corrected dissipation--coherence bound by sampling several ensembles of
finite-state continuous-time Markov jump processes.
For numerical convenience we recast the main result as a bound on the
dimensionless tightness ratio
\begin{equation}
\label{eq:tightness_ratio_def}
Y \;:=\; \frac{\Delta S}{4\pi^2\mathcal N}
\;=\; \frac{\sigma\,\lambda_R}{\lambda_I^2},
\end{equation}
so that the theorem predicts
\begin{equation}
\label{eq:tightness_ratio_bound}
Y \;\ge\; \eta.
\end{equation}

We consider three families, chosen to separate different mechanisms affecting $\eta$. The translation-invariant ring isolates symmetry-protected delocalization and serves as a reference case near $\eta=1$. The heterogeneous ring keeps the same sparse topology but introduces controlled quenched disorder, allowing one to see how localization develops as translation invariance is broken. The fully connected ensemble then shows that small $\eta$ is not merely a consequence of sparsity or near-unicyclic structure: even dense OBS-style networks can have strongly nonuniform dominant modes.

The first family is a translation-invariant biased ring.
States form a ring with nearest-neighbor transitions
$k_{i,i+1}=k_+$ and $k_{i,i-1}=k_-$ (indices mod $n$), with $k_+\neq k_-$.
This generator is circulant and its oscillatory eigenmodes are fully delocalized (detailed in Sec.~\ref{sect:eta1}), so $\eta\simeq 1$.

The second is heterogeneous biased ring.
We perturb the ring by introducing quenched disorder in the edge rates,
\begin{equation}
k_{i,i+1}=k_+\,e^{\xi_i^+},\qquad
k_{i,i-1}=k_-\,e^{\xi_i^-},
\end{equation}
where $\{\xi_i^\pm\}$ are i.i.d.\ Gaussian random variables with zero mean and variance $s^2$
(the disorder strength $s$ is varied).
This breaks translation invariance and tends to localize the oscillatory eigenmodes, reducing $\eta$.

The last one is fully connected models.
To obtain samples that are fully connected yet retain a clear dominant oscillatory mode, we build
a biased ring backbone and then add weak all-to-all links:
\begin{equation}
k_{i,i+1}=k_+,  k_{i,i-1}=k_-,
k_{ij}=k_{ji}=\varepsilon \, \text{for all other pairs},
\end{equation}
with a small $\varepsilon>0$ so the underlying graph is complete.
Finally, we introduce quenched heterogeneity by multiplying each row (outgoing rates from state $i$)
by an independent log-normal factor $e^{\zeta_i}$, $\zeta_i\sim\mathcal{N}(0,s_{\mathrm{out}}^2)$.
This preserves bidirectionality on each present edge while breaking translation invariance and producing
a broad range of mode-uniformity factors $\eta$.
Setting $\varepsilon=0$ (or switching off some links) recovers sparse topologies within the same
complete-graph parametrization, consistent with the OBS convention.

\begin{figure}[t]
 \centering
 \includegraphics[width=\linewidth]{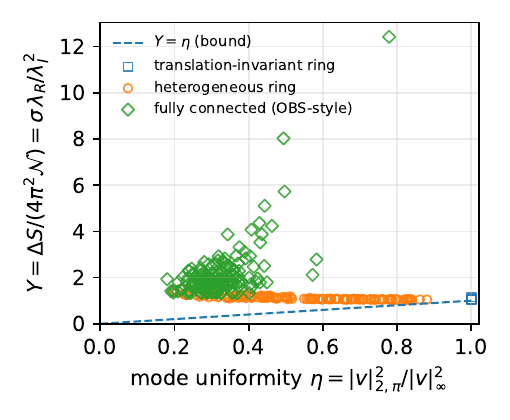}
\caption{Numerical verification of the theorem bound $Y=\sigma\lambda_R/\lambda_I^2 \ge \eta$ across model ensembles.
Hollow markers are samples from the ensembles described in Sec.~\ref{sec:numerical_example} generated with $k_+=1$ and $k_-=0.5$: translation-invariant rings with $n\in\{10,20,50,100\}$; heterogeneous rings with $n=50$ and disorder strength $s\in\{0.05,0.10,0.20,0.30\}$ (40 samples per $s$); and OBS-style fully connected models with $n=40$, all-to-all link rate $\varepsilon\in\{10^{-4},3\times10^{-4}\}$, and outgoing disorder $s_{\mathrm{out}}\in\{0.5,1.0,1.5\}$ (25 samples per $(\varepsilon,s_{\mathrm{out}})$). Only samples with coherence number $\mathcal N\ge 0.05$ are shown. The diagonal line shows $Y=\eta$.}
 \label{fig:eta_bound_scatter}
\end{figure}

Fig.~\ref{fig:eta_bound_scatter} plots the tightness ratio $Y=\sigma\lambda_R/\lambda_I^2$ versus the
mode-uniformity factor $\eta$ across all samples (open markers), together with the diagonal $Y=\eta$.
In all three ensembles we observe $Y\ge \eta$, in agreement with \eqref{eq:tightness_ratio_bound}.
Translation-invariant rings cluster near $\eta = 1$, consistent with delocalized Fourier-like modes and an OBS-like prefactor.
Introducing rate disorder on the ring lowers $\eta$ and correspondingly weakens the rigorous lower bound.
The fully connected (OBS-style) ensemble produces a wide spread in $(\eta,Y)$ at substantially smaller $\eta$,
illustrating that dense connectivity does not preclude strong localization of the dominant oscillatory mode;
nevertheless, the inequality $Y\ge\eta$ remains satisfied throughout.
We restrict the plot to samples with coherence number $\mathcal N\ge 0.05$ to focus on systems with at least weakly resolved oscillatory behavior, although the theorem itself does not require this cutoff.
Taken together, these examples show that eigenvector geometry---captured here by $\eta$---is an important ingredient controlling the sharpness of the dissipation--coherence bound beyond what is encoded in eigenvalues alone.

\section{A proof-of-principle route to estimating $\eta$ from low-dimensional data}
\label{sec:data_procedure}

\subsection{Method and scope}

In many experiments one does not observe the full Markov state $X_t$ of the underlying jump process. 
Instead one measures a stationary sampled time series
\begin{equation}
Y_0,Y_1,\dots,Y_M\in\mathbb{R}^d,
\label{eq:note_Y_series}
\end{equation}
recorded at a fixed sampling interval $\Delta t$.
Our goal in this section is deliberately modest: we outline a practically motivated route for approximating the mode-uniformity factor $\eta$ from low-dimensional data, without explicit access to the Markov generator $L$, the eigenvector $v$, or even the state $X_t$. We do \emph{not} claim a general recovery theorem. The construction below is heuristic and is intended for favorable regimes in which a single damped oscillatory mode dominates the measured dynamics over the lag window of interest.

The key point is that $\eta$ depends on how the dominant oscillatory mode is distributed, not on its overall scale.
Indeed, for the oscillatory right eigenvector $v$,
\begin{equation}
\eta
=\frac{\sum_i \pi_i |v_i|^2}{\max_i |v_i|^2}\in(0,1],
\end{equation}
so $\eta$ is a ratio of a steady-state mean-square amplitude to a peak amplitude.
If we could observe the hidden state, then along a long stationary trajectory $X_t$ the values
$v_{X_t}$ would sample the stationary weights $\pi_i$,
and one could estimate $\eta$ directly from the empirical distribution of $|v_{X_t}|$.
In experiments, $v_{X_t}$ is not observed, but it can be replaced by a surrogate mode coordinate extracted from the measured signal, provided the observations are informative enough about the dominant mode.

Concretely, we seek a complex-valued scalar coordinate $z_t$ built from $Y_t$ such that, at some lag $\tau=q\Delta t$,
it approximately evolves as a single damped oscillatory mode,
\begin{equation}
z_{t+q}\approx \hat{\mu}\,z_t,
\label{eq:note_single_mode_relation}
\end{equation}
with $\hat{\mu}$ a complex number satisfying $|\hat{\mu}|<1$ and $\arg(\hat{\mu})\neq 0$.
A convenient way to construct such a coordinate from low-dimensional measurements is to use a
(dynamic-mode/Koopman-style) linear predictor in a chosen feature space \cite{Schmid2010,Tu2014,Mezic2013,WilliamsKevrekidisRowley2015}:
pick a feature map (dictionary) $\psi:\mathbb{R}^d\to\mathbb{R}^m$ and write
\begin{equation}
\psi_t := \psi(Y_t)\in\mathbb{R}^m.
\label{eq:note_feature_vec}
\end{equation}
One then learns a lag-$\tau$ linear map in feature space and extracts an oscillatory eigenpair; the associated left eigenvector
defines a scalar projection
\begin{equation}
z_t := \hat{\xi}^\top \psi_t.
\label{eq:note_z_def}
\end{equation}
The full step-by-step construction of $\hat{\xi}$ and $\hat{\mu}$ from the time series is collected in Appendix~\ref{app:eta_recipe}.

This approach is most plausible under three conditions. First, the sampled dynamics should admit an intermediate regime in which one oscillatory mode dominates, so that a single complex number $\hat{\mu}$ captures the lagged evolution reasonably well. Second, the measured observables and chosen feature map must be rich enough that the dominant mode has a clear projection into the reconstructed coordinate. Third, the data record must be long enough, and the sampling/noise conditions favorable enough, for the amplitude statistics of $z_t$ to be stable. Outside such regimes the procedure can fail or return a surrogate that is only weakly related to the true eigenmode.

Two practical observations make this approach useful when those conditions are met.
First, $z_t$ is only defined up to an arbitrary complex rescaling (overall scale and phase):
$\hat{\xi}\mapsto c\hat{\xi}$ implies $z_t\mapsto c z_t$.
Second, $\eta$ is scale invariant, so any such rescaling cancels out in the estimator below.
Thus, once a meaningful single-mode coordinate $z_t$ has been extracted, one can estimate $\eta$ from the stationary
amplitude statistics of $|z_t|$ alone.

Given $z_t$, we use the scale-invariant plug-in estimator
\begin{equation}
\widehat{\eta}:=\frac{\frac{1}{N_{\mathrm{pair}}}\sum_{t=0}^{N_{\mathrm{pair}}-1}|z_t|^2}{\left(\max_{0\le t\le N_{\mathrm{pair}}-1}|z_t|\right)^2},
\label{eq:note_eta_hat}
\end{equation}
where $N_{\mathrm{pair}}=M-q$ is the number of lagged pairs.
In finite data, the sample maximum can be sensitive to rare spikes or outliers.
A robust alternative replaces the sample maximum by a high quantile of $|z_t|^2$,
\begin{equation}
\widehat{\eta}_{\mathrm{rob}}
:=\frac{\frac{1}{N_{\mathrm{pair}}}\sum_{t=0}^{N_{\mathrm{pair}}-1}|z_t|^2}{Q_{1-\varepsilon}\!\left(|z|^2\right)},
\label{eq:note_eta_hat_rob}
\end{equation}
where $Q_{1-\varepsilon}(|z|^2)$ is the empirical $(1-\varepsilon)$-quantile of $\{|z_t|^2\}$.
Because no general consistency theorem is proved here, it is good practice to test the stability of $\widehat{\eta}$ under changes of lag, dictionary, and subsampling.

Finally, with multiple measured observables one can alternatively fit the cross-correlation matrix to a single damped oscillatory mode
\cite{OhgaItoKolchinsky2023,Shiraishi2023} and (when observables are sufficiently informative) reconstruct the oscillatory mode up to scale and phase.
This route is even more restrictive: it requires a common single-mode window across observables and measurements informative enough to reconstruct the mode. We present it in Appendix~\ref{app:corr} only as an optional consistency check.

 \subsection{Example}
\label{sec:demo5_nonunicyclic_PRE}

We illustrate the practical procedure on a five-state continuous-time Markov jump process, following Appendix~\ref{app:eta_recipe}. The state space is $\{1,2,3,4,5\}$. The network consists of a biased nearest-neighbor ring together with an additional chord connecting states $1$ and $3$, so that the undirected edge set has $E=6$ edges and the cycle rank is $E-n+1=2$. Equivalently, the network contains two independent cycles, namely $1\to2\to3\to1$ and $1\to3\to4\to5\to1$.

On the ring, we set $k_{i,i+1}=a_i$ and $k_{i,i-1}=b_i$ with indices understood modulo $5$. The parameters are chosen as
\begin{equation}
(a_1,a_2,a_3,a_4,a_5)=(0.5,\,4,\,4,\,4,\,4),\qquad
 b_i=1,
\end{equation}
and the chord rates are
\begin{equation}
k_{13}=k_{31}=2.
\end{equation}
The forward generator $W$ is defined as in Eq.~\eqref{eq:generator} (with column-vector convention), and the stationary distribution $\pi$ satisfies $W\pi=0$ and $\sum_i \pi_i=1$. We set $L:=W^{\top}$ and compute the dominant oscillatory eigenpair $Lv=\lambda v$ by selecting the eigenvalue with $\text{Im}(\lambda)>0$ and $\text{Re}(\lambda)$ closest to $0$. For the parameters above one obtains
\begin{equation}
\lambda_{\mathrm{true}}\approx -4.4863238 + 1.7825632\,\mathrm{i}.
\end{equation}
The associated mode-uniformity factor in Eq.~\eqref{eq:eta_def} evaluates to
\begin{equation}
\eta_{\mathrm{true}}\approx 0.6045616.
\end{equation}

This example is intentionally favorable: the observation noise is weak, the trajectory is long, and the dictionary is centered at the noiseless observation points. It is therefore meant as a proof of principle that the construction can succeed under near-ideal conditions, not as a systematic robustness benchmark against shorter time series, poorer feature choices, coarser sampling, or multimode contamination.

We next demonstrate that $\eta$ can be recovered from low-dimensional observations. We do not observe $X_t$ directly; instead we observe $Y_t\in\mathbb{R}^2$ given by
\begin{equation}
\begin{aligned}
&Y_t=h(X_t)+\sigma_{\mathrm{obs}}\varepsilon_t,\\
&h(i)=\big(\cos(2\pi(i-1)/5),\,\sin(2\pi(i-1)/5)\big),	
\end{aligned}
\end{equation}
where $\varepsilon_t\sim\mathcal N(0,I_2)$ are i.i.d.\ and $\sigma_{\mathrm{obs}}=0.01$. A stationary trajectory is sampled at interval $\Delta t=0.1$, producing $Y_0,\dots,Y_M$ with $M=200{,}000$ using random seed $1$. We take $q=1$, hence $\tau=q\Delta t=0.1$.

Following Sec.~\ref{sec:data_procedure}, we use an RBF dictionary centered at the five noiseless observation points $c_j=h(j)$,
\begin{equation}
\psi_j(y)=\exp\!\Big(-\frac{\|y-c_j\|^2}{2\sigma_{\mathrm{rbf}}^2}\Big),
\qquad \sigma_{\mathrm{rbf}}=0.4,
\end{equation}
so that $\psi(y)\in\mathbb{R}^m$ with $m=5$. Writing $\psi_t=\psi(Y_t)$ and forming the snapshot matrices $\Psi=[\psi_0,\dots,\psi_{M-q-1}]$ and $\Psi'=[\psi_q,\dots,\psi_{M-1}]$, we compute
\begin{equation}
\widehat C_{00}=\frac{1}{M-q}\Psi\Psi^{\top}, \quad 
\widehat C_{0\tau}=\frac{1}{M-q}\Psi(\Psi')^{\top}
\end{equation}
and
\begin{equation}
	\widehat T_{\tau}=(\widehat C_{00}+\beta I)^{-1}\widehat C_{0\tau},
\end{equation}
with ridge parameter $\beta=10^{-8}$. We then compute a left eigenpair $\widehat T_{\tau}^{\top}\hat\xi=\hat\mu\,\hat\xi$, select $\hat\mu$ by maximizing the coherence proxy $\widehat{\mathcal N}(\hat\mu)$ in Eq.~\eqref{eq:note_coherence_proxy} among eigenvalues with $|\hat\mu|<1$ and $\arg(\hat\mu)\in(0,\pi)$, and define the scalar mode coordinate $z_t=\hat\xi^{\top}\psi_t$. 
Finally, the practical estimator of $\eta$ is
\begin{equation}
\widehat\eta\approx 0.6046444,
\end{equation}
which shows that the method of Sec.~\ref{sec:data_procedure} can recover $\eta$ accurately from low-dimensional noisy observations in a non-unicyclic five-state network under these favorable conditions. A broader robustness study is beyond the scope of the present work.

\section{An eigenvector-free bound}
\label{sect:piminbound}

When reconstructing $\eta$ is infeasible, one can still obtain the following rigorous bound using only the smallest stationary weight $\pi_{\min}:=\min_i \pi_i$, which is often accessible experimentally.

This corollary follows from the inequality $\eta\ge \pi_{\min}$, which is an immediate consequence of the definition \eqref{eq:eta_def}.
For completeness, let $i^\star$ satisfy $|v_{i^\star}|^2=\max_i|v_i|^2=\norminf{v}^2$. Then
\[
\eta=\frac{\sum_i\pi_i|v_i|^2}{|v_{i^\star}|^2}
=\sum_i\pi_i\frac{|v_i|^2}{|v_{i^\star}|^2}
\ge \pi_{i^\star}\ge \pi_{\min}.
\]
Substituting $\eta\ge \pi_{\min}$ into Eqs.~\eqref{eq:MainSigma} and \eqref{eq:MainDeltaS_proved} yields the eigenvector-free bound
\begin{equation}
\Delta S \ge 4\pi^{2} \pi_{\min}\,\mathcal{N}.
\label{eq:piMinCor}
\end{equation}
It is typically weaker than a direct estimate of $\eta$, but is robust and requires minimal structural inference.

\section{When does the correction disappear? The case \texorpdfstring{$\eta=1$}{eta=1}}

\label{sect:eta1}

The eigenvector-free bound \eqref{eq:piMinCor} is robust but may be weak when $\pi_{\min}$ is small. At the opposite extreme, the mode-uniformity factor can attain its maximum value $\eta=1$. This case has a simple interpretation. Because
\[
\eta=\frac{\sum_i \pi_i |v_i|^2}{\max_i |v_i|^2}\le 1,
\]
equality holds if and only if the oscillatory eigenmode has constant modulus on all states with positive stationary weight. Thus $\eta=1$ is exactly the condition of complete delocalization in the steady-state metric. Translation invariance on a ring enforces this structure: circulant generators are diagonalized by Fourier waves, and every nontrivial Fourier mode has constant modulus. For this symmetry-protected class the localization correction disappears, so the refined bound collapses to the OBS form. The saturation result discussed below is therefore not separate from the main theorem; it identifies a concrete mechanism by which the original OBS prefactor is recovered.

\subsection{Translation-invariant jump processes on a ring}
\label{sec:LambdaRing}

Let the state space be the cyclic group $\mathbb{Z}_{n}=\{0,1,\dots,n-1\}$ with addition modulo $n$, and label states by $x\in\mathbb{Z}_{n}$.
A translation-invariant continuous-time random walk is specified by rates $k_{r}\ge 0$ for each nonzero displacement $r\in\{1,2,\dots,n-1\}$, meaning that the process jumps
\[
x\longrightarrow x+r\ (\mathrm{mod}\ n)\quad\text{at rate }k_{r}.
\]
The corresponding generator is circulant, with entries
\begin{equation}
R_{x+r,x}=k_{r}\quad (r=1,\dots,n-1),
\qquad
R_{x,x}=-\sum_{r=1}^{n-1}k_{r},
\label{eq:ring_general_R}
\end{equation}
and $R_{y,x}=0$ otherwise (indices understood modulo $n$).
Translation invariance implies a uniform steady state, $\pi_x=1/n$.

Fourier modes diagonalize $R$ \cite{Diaconis1988,LevinPeresWilmer2009}. For each $m\in\{0,1,\dots,n-1\}$ define $q_m=2\pi m/n$ and
\begin{equation}
u^{(m)}_{x}=\exp(-\ii q_m x).
\label{eq:ring_fourier_mode}
\end{equation}
Substituting into $Ru=\lambda u$ gives
\begin{equation}
\lambda_m
=
\sum_{r=1}^{n-1}k_{r}\bigl(\ee^{\ii q_m r}-1\bigr)
\equiv -\gamma_m+\ii\omega_m,
\label{eq:ring_general_lambda}
\end{equation}
with real and imaginary parts
\begin{align}
\gamma_m
&=
\sum_{r=1}^{n-1}k_{r}\bigl(1-\cos(q_m r)\bigr),
\label{eq:ring_general_gamma}\\
\omega_m
&=
\sum_{r=1}^{n-1}k_{r}\sin(q_m r).
\label{eq:ring_general_omega}
\end{align}
(For $m$ and $n-m$ the eigenvalues are complex conjugates.)

Because $|u^{(m)}_{x}|=1$ for all $x$ and all $m\neq 0$, the oscillatory mode has constant modulus in the steady-state metric and therefore
\begin{equation}
\eta=1
\qquad\text{for every nontrivial Fourier mode.}
\label{eq:ring_general_eta1}
\end{equation}

The steady-state entropy production rate for this ring can be written by pairing each displacement $r$ with its reverse $n-r$:
\begin{equation}
\sigma
=
\sum_{r=1}^{\lfloor (n-1)/2\rfloor}\bigl(k_{r}-k_{n-r}\bigr)\ln\frac{k_{r}}{k_{n-r}}.
\label{eq:ring_general_sigma}
\end{equation}
If $n$ is even, the self-inverse displacement $r=n/2$ contributes nothing to $\sigma$ because $n-r=r$.

\subsection{Nearest-neighbor biased ring}

A commonly used unicyclic model is the nearest-neighbor ring, recovered by taking only the displacements $r=1$ and $r=n-1$ nonzero:
\begin{equation}
k_{1}=k^{+},\qquad k_{n-1}=k^{-},\qquad k_{r}=0\ \text{for }r\notin\{1,n-1\},
\label{eq:ring_nn_rates}
\end{equation}
with $k^{+}\neq k^{-}$ to break detailed balance. Eq.~\eqref{eq:ring_general_lambda} reduces to
\begin{align}
\lambda_m
&=
-(k^{+}+k^{-}) + k^{+}\ee^{2\pi\ii m/n}+k^{-}\ee^{-2\pi\ii m/n}
\nonumber\\
&=
-(k^{+}+k^{-})\bigl(1-\cos\theta_m\bigr) + \ii (k^{+}-k^{-})\sin\theta_m,
\label{eq:ringEig}
\end{align}
where $\theta_m=2\pi m/n$.
For a given oscillatory eigenvalue $\lambda_m=-\gamma_m+\ii\omega_m$ one has
\begin{equation}
\gamma_m=(k^{+}+k^{-})\bigl(1-\cos\theta_m\bigr),
\qquad
\omega_m=(k^{+}-k^{-})\sin\theta_m.
\label{eq:ring_gw}
\end{equation}
The coherence number $\mathcal{N}_m=\omega_m/(2\pi\gamma_m)$ is
\begin{equation}
\mathcal{N}_m
=
\frac{k^{+}-k^{-}}{2\pi(k^{+}+k^{-})} \frac{\sin\theta_m}{1-\cos\theta_m}
=
\frac{k^{+}-k^{-}}{2\pi(k^{+}+k^{-})} \cot\frac{\theta_m}{2}.
\label{eq:ringN}
\end{equation}
For the slowest oscillatory mode $m=1$ on a large ring, $\theta_1=2\pi/n$ and $\cot(\theta_1/2)\simeq n/\pi$, so $\mathcal{N}_1$ grows linearly with $n$ at fixed bias ratio $k^{+}/k^{-}$.

\subsection{Saturation: diffusive continuum limit on the ring}
\label{sec:diffusion_limit}

The prefactor $4\pi^{2}$ obtained when $\eta=1$ is not merely an artifact of our inequalities: it is optimal even within the fully delocalized, translation-invariant class. This is seen most cleanly in the diffusive continuum limit, where a biased random walk on an increasingly fine ring converges to drift--diffusion on a circle and our bound becomes an equality.

Consider drift--diffusion (Fokker--Planck) dynamics on the circle $x\in[0,2\pi)$ \cite{Risken1989,Gardiner2009},
\begin{equation}
\partial_t \rho(x,t) = -v_{\mathrm{dr}} \partial_x \rho(x,t) + D \partial_x^{2}\rho(x,t),
\qquad v_{\mathrm{dr}}\in\mathbb{R},\ D>0.
\label{eq:driftdiff}
\end{equation}
The steady state is uniform. Fourier modes $\ee^{\ii n x}$ are eigenfunctions with eigenvalues
\begin{equation}
\lambda_n = -D n^{2} + \ii v_{\mathrm{dr}} n,\qquad n\in\mathbb{Z}.
\label{eq:diff_eigs}
\end{equation}
For the fundamental oscillatory mode $n=1$ one has $\lambda_R=D$ and $\lambda_I=v_{\mathrm{dr}}$, so $\mathcal{N}=v_{\mathrm{dr}}/(2\pi D)$ and the period is $T=2\pi/v_{\mathrm{dr}}$.

The steady-state probability current is constant, $J=v_{\mathrm{dr}}/(2\pi)$, and the entropy production rate is \cite{Seifert2012,Risken1989}
\begin{equation}
\sigma = \frac{v_{\mathrm{dr}}^{2}}{D}.
\label{eq:sigma_diff}
\end{equation}
Therefore
\begin{equation}
\sigma = \frac{\lambda_I^{2}}{\lambda_R},
\qquad
\Delta S = \sigma \frac{2\pi}{\lambda_I} = 4\pi^{2}\mathcal{N},
\end{equation}
so the OBS prefactor is achieved with equality in this limit.

A nearest-neighbor biased ring approaches Eq.~\eqref{eq:driftdiff} under the standard diffusive scaling. Taking lattice spacing $a=2\pi/n$ and rates
\begin{equation}
k^{+}=D/a^{2}+v_{\mathrm{dr}}/(2a),\qquad k^{-}=D/a^{2}-v_{\mathrm{dr}}/(2a),
\label{eq:diffusive_scaling_rates}
\end{equation}
(with $n$ large enough that both rates are positive), the fundamental Fourier eigenvalue tends to $-D+\ii v_{\mathrm{dr}}$ as $n\to\infty$. In this limit the inequality $\sigma\ge \eta\,\lambda_I^{2}/\lambda_R$ with $\eta=1$ becomes arbitrarily close to equality, illustrating the tightness of the delocalized-mode bound.

\section{Relation to existing thermodynamic bounds on oscillations}

Thermodynamic aspects of stochastic oscillations have been studied from several complementary angles. In biochemical oscillator models, free-energy dissipation has been related to phase accuracy, phase sensitivity, synchronization, and design under limited energetic resources \cite{CaoWangOuyangTu2015,FeiCaoOuyangTu2018,ZhangCaoOuyangTu2020,CaoJiangHou2020,MarslandCuiHorowitz2019}. At the broader level of reaction-network thermodynamics, stochastic thermodynamics provides a general framework for open chemical reaction networks and their nonequilibrium driving \cite{Seifert2012,RaoEsposito2016}. These works establish important physical context for the present problem, but they typically focus on particular oscillator architectures, on thermodynamic uncertainty-type comparisons, or on performance measures other than the dominant spectral coherence number used here.

In Markov-network terms, a number of general inequalities constrain oscillatory behavior, but they differ in the oscillation metric being constrained, the thermodynamic quantities that enter, and the extent to which they depend only on eigenvalues versus also on eigenvectors and observables \cite{Seifert2012}. Early results connected coherence measures such as the coherence number to nonequilibrium driving strength and network structure, with explicit statements for unicyclic networks and conjectures for broader classes \cite{BaratoSeifert2017}. Subsequent work derived general inequalities for the asymmetry of two-time cross-correlations controlled by thermodynamic driving, and used these to establish coherence bounds that had been conjectured previously \cite{OhgaItoKolchinsky2023,LiangPigolotti2023}. Shiraishi derived limits on oscillatory behavior in correlation functions and fluctuations controlled by entropy production per characteristic timescale \cite{Shiraishi2023}. Gu obtained bounds on normalized correlation asymmetry in terms of entropy production together with dynamical activity \cite{Gu2024}. Kolchinsky, Ohga, and Ito compared a nonequilibrium generator with a reference equilibrium generator and derived spectral bounds with consequences for relaxation and the imaginary parts of eigenvalues \cite{KolchinskyOhgaIto2024}. These results are powerful but either bound coherence at fixed driving, concern oscillatory signatures of chosen observables, or introduce an explicit reference dynamics.

Our inequality instead directly lower-bounds the entropy produced per oscillation period in terms of the dominant oscillatory eigenpair, in the same spirit as the conjecture posed by Oberreiter, Barato, and Seifert \cite{OBS2022}. The new point is that the sharp prefactor is not controlled by eigenvalues alone: it also depends on how delocalized the corresponding eigenvector is in the steady-state inner product. The factor $\eta$ makes this dependence explicit. When symmetry forces delocalization, as on translation-invariant rings, the correction disappears and the refined bound reduces to the OBS form. When localization is allowed, the theorem gives a controlled weakening rather than an unconditional universal prefactor. In this sense the present result identifies a concrete criterion under which eigenvalue-only dissipation--coherence bounds become sharp.

\section{Conclusion}

We established a rigorous dissipation--coherence inequality for finite nonequilibrium Markov jump processes with a dominant oscillatory relaxation mode. The new ingredient is the mode-uniformity factor, which captures how delocalized the oscillatory eigenvector is in the steady-state inner product. This exposes a concrete mechanism by which an eigenvalue-only universal prefactor can fail: even at fixed spectral coherence, localized oscillatory modes can substantially reduce the minimum dissipation implied by thermodynamics.

Beyond the theorem itself, we proposed two routes that make the bound operational at different levels of information. First, under single-mode dominance and sufficiently informative low-dimensional measurements, a surrogate oscillatory coordinate can in favorable cases be used to approximate $\eta$ from stationary amplitude statistics. The present data-driven section should therefore be read as a proof of principle rather than as a general inference theorem. Second, when eigenvector information is unavailable, a conservative eigenvector-free corollary follows from $\eta\ge\pi_{\min}$, yielding $\Delta S\ge 4\pi^2\pi_{\min}\mathcal{N}$. These complementary statements clarify what can be inferred from data of increasing richness.

We also identified an important delocalized class for which $\eta=1$ and the OBS prefactor $4\pi^2$ is recovered as a rigorous bound: translation-invariant jump processes on a ring (circulant generators) \cite{Diaconis1988,LevinPeresWilmer2009}. In the drift--diffusion limit on a circle, the inequality becomes an equality, showing that the $4\pi^2$ prefactor is tight in this symmetry-protected setting \cite{Risken1989,Seifert2012}. This symmetry-based characterization explains how the refinement connects back to the original OBS conjecture: whenever the dominant oscillatory mode is forced to be fully delocalized, the localization correction disappears.

Several directions follow. A first goal is to characterize, for physically motivated network ensembles and biochemical models, when the dominant oscillatory mode is delocalized (so $\eta\simeq 1$) and how $\eta$ scales with system size and heterogeneity. A second is to turn the present proof-of-principle inference route into a systematic estimator with principled uncertainty quantification, robustness tests under shorter and noisier observations, and explicit treatment of multi-mode contamination \cite{OhgaItoKolchinsky2023,Shiraishi2023}. Finally, it would be interesting to extend the localization-corrected tradeoff to coarse-grained descriptions and to settings beyond finite-state continuous-time jumps \cite{santolin2025,kolchinsky2025,nagayama2025a}, such as diffusion processes, periodically driven Markov dynamics, or generators with multiple competing oscillatory modes.

\appendix

\section{Proof of the main theorem}
\label{app:proof}

We recall the definitions from Sec.~II: stationary flows $a_{ij}=\pi_i k_{ij}$, currents $J_{ij}=a_{ij}-a_{ji}$, and traffic $T_{ij}=a_{ij}+a_{ji}$, so that $\sigma$ is given by Eq.~\eqref{eq:sigma_def}.

The scalar inequality $(x-y)\ln(x/y)\ge 2(x-y)^2/(x+y)$ for $x,y>0$ applied to $(x,y)=(a_{ij},a_{ji})$ yields \cite{Shiraishi2018,Shiraishi2021a,Dechant2022a}
\begin{equation}
J_{ij}A_{ij} \ge 2\frac{J_{ij}^2}{T_{ij}},
\label{eq:edge_current_bound}
\end{equation}
and summing over $i<j$ gives
\begin{equation}
\EP \ge 2\sum_{i<j}\frac{J_{ij}^2}{T_{ij}}.
\label{eq:global_current_bound}
\end{equation}

Let $L^\dagger$ denote the adjoint of $L$ with respect to $\ip{\cdot}{\cdot}$.
Define $L_s=(L+L^\dagger)/2$ and $L_a=(L-L^\dagger)/2$.
Then $L_s$ is self-adjoint and $L_a$ is anti-self-adjoint in $\ip{\cdot}{\cdot}$.

Define the Dirichlet form \cite{LevinPeresWilmer2009}
\begin{equation}
\Dform(u) := \frac{1}{2}\sum_{i<j} T_{ij}\,|u_i-u_j|^2.
\label{eq:Dirichlet}
\end{equation}
One has the standard quadratic forms
\begin{equation}
-\ip{u}{L_s u}=\Dform(u),
\qquad
\ip{u}{L_a u}=\ii \sum_{i<j} J_{ij}\,\mathrm{Im}(u_i^* u_j).
\label{eq:forms}
\end{equation}
For the eigenvector $v$ in Eq.~\eqref{eq:eigenpair}, taking $\ip{v}{\cdot}$ yields
\begin{equation}
\lambda_R = \frac{\Dform(v)}{\normpi{v}^2},
\qquad
\lambda_I = \frac{1}{\normpi{v}^2}\sum_{i<j} J_{ij}\,\mathrm{Im}(v_i^* v_j).
\label{eq:lambdas_forms}
\end{equation}

Start from Eq.~\eqref{eq:lambdas_forms} and apply Cauchy--Schwarz with weights $T_{ij}$:
\begin{equation}
\left(\sum_{i<j} J_{ij}\mathrm{Im}(v_i^* v_j)\right)^2
\le
\left(\sum_{i<j}\frac{J_{ij}^2}{T_{ij}}\right)
\left(\sum_{i<j}T_{ij}\,[\mathrm{Im}(v_i^* v_j)]^2\right).
\label{eq:CS1}
\end{equation}
Using Eq.~\eqref{eq:global_current_bound}, $\sum_{i<j} J_{ij}^2/T_{ij}\le \EP/2$, so
\begin{equation}
\lambda_I^2 \le \frac{\EP}{2}\,\frac{1}{\normpi{v}^4}
\left(\sum_{i<j}T_{ij}\,[\mathrm{Im}(v_i^* v_j)]^2\right).
\label{eq:lambdaI_mid}
\end{equation}

Next, the two-point complex inequality
\begin{equation}
[\mathrm{Im}(a^*b)]^2 \le \frac{1}{2}\left(|a|^2+|b|^2\right)|a-b|^2
\label{eq:complex_two_point}
\end{equation}
implies, after multiplying by $T_{ij}$ and summing over $i<j$,
\begin{equation}
\sum_{i<j}T_{ij}\,[\mathrm{Im}(v_i^* v_j)]^2
\le
\norminf{v}^2 \sum_{i<j}T_{ij}|v_i-v_j|^2
=
2\,\norminf{v}^2\,\Dform(v).
\label{eq:Im_bound}
\end{equation}
Substituting Eq.~\eqref{eq:Im_bound} into Eq.~\eqref{eq:lambdaI_mid} and using $\Dform(v)=\lambda_R\normpi{v}^2$ yields
\begin{equation}
\lambda_I^2 \le \frac{\norminf{v}^2}{\normpi{v}^2}\,\EP\,\lambda_R.
\label{eq:lambdaI_ratio}
\end{equation}
Rearranging and using $\eta=\normpi{v}^2/\norminf{v}^2$ from Eq.~\eqref{eq:eta_def} gives Eq.~\eqref{eq:MainSigma}. Equation~\eqref{eq:MainDeltaS_proved} follows by multiplying by $2\pi/\lambda_I$ and using $\mathcal{N}=\lambda_I/(2\pi\lambda_R)$.

\section{Recipe for estimating $\eta$ from low-dimensional time series}
\label{app:eta_recipe}

This appendix collects an implementation-oriented version of the construction summarized in Sec.~\ref{sec:data_procedure}. It should be read as a heuristic recipe rather than a theorem guaranteeing recovery of $\eta$. The procedure is most plausible when the sampled dynamics is well approximated by a single damped oscillatory mode at the chosen lag and when the feature map resolves that mode. We assume a stationary time series $\{Y_t\}_{t=0}^M$ sampled at interval $\Delta t$ as in Eq.~\eqref{eq:note_Y_series}.

\begin{enumerate}
\item \textbf{Choose a lag.}
Pick an integer $q\ge 1$ and set the lag $\tau=q\Delta t$.
Form snapshot pairs $(Y_t,Y_{t+q})$ for $t=0,1,\dots,M-q-1$, and set $N_{\mathrm{pair}}:=M-q$.

\item \textbf{Choose a feature map.}
Pick a dictionary $\psi:\mathbb{R}^d\to\mathbb{R}^m$ (linear features $\psi(y)=y$ reduce to standard DMD \cite{Schmid2010,Tu2014}).
Define $\psi_t=\psi(Y_t)$ as in Eq.~\eqref{eq:note_feature_vec} and build the snapshot matrices
\begin{equation}
\begin{aligned}
	&\Psi := \big[\psi_0,\psi_1,\dots,\psi_{N_{\mathrm{pair}}-1}\big]\in\mathbb{R}^{m\times N_{\mathrm{pair}}},
\\
&\Psi' := \big[\psi_q,\psi_{q+1},\dots,\psi_{q+N_{\mathrm{pair}}-1}\big]\in\mathbb{R}^{m\times N_{\mathrm{pair}}}.
\end{aligned}
\label{eq:note_Psi_mats}
\end{equation}

\item \textbf{Fit a lag-$\tau$ linear predictor in feature space \cite{WilliamsKevrekidisRowley2015}.}
Compute empirical covariances
\begin{equation}
\widehat{C}_{00}:=\frac{1}{N_{\mathrm{pair}}}\Psi\Psi^\top,
\qquad
\widehat{C}_{0\tau}:=\frac{1}{N_{\mathrm{pair}}}\Psi(\Psi')^\top,
\label{eq:note_covs}
\end{equation}
choose a ridge parameter $\beta\ge 0$ (useful when $\widehat{C}_{00}$ is ill-conditioned), and set
\begin{equation}
\widehat{T}_\tau := \big(\widehat{C}_{00}+\beta I\big)^{-1}\widehat{C}_{0\tau}.
\label{eq:note_T_hat}
\end{equation}

\item \textbf{Extract an oscillatory eigenpair.}
Compute a left eigenpair of $\widehat{T}_\tau$,
\begin{equation}
\widehat{T}_\tau^T \hat{\xi}=\hat{\mu}\,\hat{\xi}.
\label{eq:note_eig}
\end{equation}
Select a complex eigenvalue $\hat{\mu}$ with $|\hat{\mu}|<1$ and $\arg(\hat{\mu})\in(0,\pi)$ that represents a decaying oscillation.
A convenient selection criterion is to maximize the discrete-time coherence proxy
\begin{equation}
\widehat{\mathcal{N}}(\hat{\mu}) := -\frac{1}{2\pi}\frac{\arg(\hat{\mu})}{\log|\hat{\mu}|},
\label{eq:note_coherence_proxy}
\end{equation}
subject to $\arg(\hat{\mu})\neq 0$ and $|\hat{\mu}|<1$.

If desired, convert $\hat{\mu}$ into continuous-time parameters via
\begin{equation}
\hat{\lambda}=\frac{1}{\tau}\log \hat{\mu},
\qquad
\hat{\lambda}_R=-\mathrm{Re}(\hat{\lambda}),
\qquad
\hat{\lambda}_I=\mathrm{Im}(\hat{\lambda}),
\label{eq:note_lambda_from_mu}
\end{equation}
choosing the branch of the logarithm to match the observed oscillation frequency.

\item \textbf{Form the scalar mode coordinate and diagnose single-mode dominance.}
Define $z_t=\hat{\xi}^\top\psi_t$ as in Eq.~\eqref{eq:note_z_def}. In practice one should verify that the single-mode relation $z_{t+q}\approx \hat{\mu}z_t$ is at least approximately satisfied over the time window of interest.

\item \textbf{Estimate $\eta$.}
Compute $\widehat{\eta}$ from Eq.~\eqref{eq:note_eta_hat} (or the robust variant $\widehat{\eta}_{\mathrm{rob}}$ in Eq.~\eqref{eq:note_eta_hat_rob}). Because no general consistency theorem is proved here, it is good practice to check stability under changes of $q$, of the feature dictionary, and of subsampling or bootstrap resampling.
\end{enumerate}

\section{Correlation-based estimation of $\eta$}
\label{app:corr}

In many experiments one cannot directly observe the full Markov state $X_t$. Instead one records a small set of physical observables (fluorescence signals, FRET efficiencies, positions, etc.) and studies their time correlations. Cross-correlation based approaches can already constrain nonequilibrium behavior and oscillations \cite{OhgaItoKolchinsky2023,Shiraishi2023}. Here we highlight a complementary viewpoint. This route is more restrictive than the surrogate-coordinate method of Appendix~\ref{app:eta_recipe}: it requires a time window dominated by a single oscillatory relaxation mode across all measured observables and, for actual reconstruction of $v$, enough independent observables to resolve that mode. Under those conditions, the oscillatory part of the measured correlations carries information about how that mode projects onto the chosen observables. With sufficiently many independent observables, one can (in principle) reconstruct the eigenmode up to an overall scale and phase, and thereby estimate the quantities $|v_i|^2$ that enter $\eta$.

Choose $m$ experimentally accessible observables $f^{(\alpha)}(i)$, $\alpha=1,\dots,m$, defined on the discrete states $i=1,\dots,n$.
From a stationary time series, form the centered signals
\[
\delta f^{(\alpha)}(t)=f^{(\alpha)}(X_t)-\langle f^{(\alpha)}\rangle_\pi,
\qquad
\langle f^{(\alpha)}\rangle_\pi=\sum_i \pi_i f^{(\alpha)}(i),
\]
and estimate the stationary cross-correlation matrix
\begin{equation}
C_{\alpha\beta}(t)=\langle \delta f^{(\alpha)}(t)\,\delta f^{(\beta)}(0)\rangle_\pi.
\end{equation}

If a single oscillatory eigenmode $\lambda=-\lambda_R+\ii\lambda_I$ dominates the dynamics over some intermediate time window (after fast transients have died out but before the signal is lost in noise), then all entries of $C_{\alpha\beta}(t)$ share the same decay rate $\lambda_R$ and frequency $\lambda_I$ \cite{VanKampen2007,Gardiner2009}. In that regime one expects the approximate form
\begin{equation}
C_{\alpha\beta}(t)\approx
\ee^{-\lambda_R t}\Bigl[
M_{\alpha\beta}\cos(\lambda_I t)+N_{\alpha\beta}\sin(\lambda_I t)
\Bigr],
\label{eq:C_fit}
\end{equation}
where $M$ and $N$ are constant (time-independent) amplitude matrices.
A practical procedure is to fit Eq.~\eqref{eq:C_fit} simultaneously over all pairs $(\alpha,\beta)$ to extract
$(\widehat{\lambda}_R,\widehat{\lambda}_I)$ and the fitted matrices $(\widehat M,\widehat N)$.
It is convenient to combine them into a complex amplitude matrix
\begin{equation}
\widehat A_{\alpha\beta}:=\widehat M_{\alpha\beta}-\ii\,\widehat N_{\alpha\beta}.
\end{equation}

Under the single-mode approximation, the oscillatory contribution to correlations factorizes into an ``input'' part and an ``output'' part, so the matrix $\widehat A$ is approximately rank one:
\[
\widehat A \approx \bm{p}\,\bm{q}^{\mathsf T}.
\]
In practice one can extract $\bm{p}$ and $\bm{q}$ by taking the leading singular vectors of $\widehat A$, i.e., keeping only the largest singular value in an SVD.

The vectors $\bm{p}$ and $\bm{q}$ live in the $m$-dimensional observable space. To relate them to the $n$-dimensional eigenvector $v$ of $L$, introduce the $m\times n$ matrix of steady-state weighted (and centered) observables,
\begin{equation}
F_{\alpha i}:=\sqrt{\pi_i}\,\delta f^{(\alpha)}(i).
\label{eq:Fmat}
\end{equation}
In the idealized case $m\ge n$ where the columns of $F$ are linearly independent, one can reconstruct the oscillatory eigenvector up to a scalar factor by a least-squares inversion:
\begin{equation}
\widehat{\bm v}\ \propto\ F^{+}\,\widehat{\bm p},
\end{equation}
where $F^{+}$ is the Moore--Penrose pseudoinverse and $\widehat{\bm p}$ is the leading left singular vector from the rank-one approximation of $\widehat A$.
After fixing the arbitrary complex phase (e.g.\ by choosing a reference component to be real and positive) and normalizing by $\max_i|\widehat v_i|^2=1$, one computes
$\widehat\eta=\sum_i \pi_i|\widehat v_i|^2$ in analogy with Eq.~\eqref{eq:eta_def}.

When fewer observables are available than states, $F$ cannot have full column rank. The same construction then yields only a projection of $v$ onto the span of the measured observables. In this case the resulting $\widehat\eta$ should be interpreted as an effective mode-uniformity estimate for the chosen measurement subspace. If such reconstruction is unreliable, the conservative eigenvector-free bound based on $\pi_{\min}$ remains available as a robust fallback.

\bibliography{ref}

\end{document}